\documentclass[a4paper]{article}

\usepackage[english]{babel}
\usepackage[utf8x]{inputenc}
\usepackage[T1]{fontenc}

\usepackage[a4paper,top=3cm,bottom=2cm,left=3cm,right=3cm,marginparwidth=1.75cm]{geometry}

\usepackage{amsmath}
\usepackage{graphicx}
\usepackage[colorinlistoftodos]{todonotes}
\usepackage[colorlinks=true, allcolors=blue]{hyperref}
\usepackage{authblk}
\usepackage[toc,page]{appendix}

\title{Black Hole Complementarity and Violation of Causality}
\author{Moshe Rozenblit\thanks{Moshe\_Rozenblit@yahoo.com}}
\affil{New York Quantum Theory Group}

\begin{document}
\maketitle

\begin{abstract}
Analysis of a massive shell collapsing on a solid sphere shows that black hole complementarity (BHC) violates causality in its effort to save information conservation. In particular, this note describes a hypothetical contraption based on BHC that would allow the transfer of information from the future to the present. 

\end{abstract}

\section{Introduction} 

Black hole (BH) complementarity \cite{1} has been the most popular vehicle deployed to address the BH information paradox. However, under some circumstances, application of that paradigm to the formation of a BH can lead to the violation of causality. Section 2 examines a massive shell collapsing on a billiard ball to form a BH and, in particular, how BH complementarity (BHC) preserves the information embedded in the billiard ball. It also argues, however, that BHC leads to the violation of causality. Section 3 goes further as it analyses the case of a shell collapsing on a giant massive solid sphere and shows how BHC allows for the transmission of one bit of information from the future to the present. An Appendix provides a more picturesque yet more compelling view of backwards transmission of information and generalizes the procedure to the transmission of any number of bits from the future to the present, thus making BHC’s causality violation manifestly explicit.

\section{Massive shell collapsing on a billiard ball}

Consider a spherically symmetric, uniform and rigid thin shell, with zero charge, zero angular momentum, $100\%$ transparent at all wavelengths and mass M such that $2GM/c2 = 1$ light year (l.y.) where G is Newton’s constant and c is the speed of light in vacuum. The shell’s radius is $1 l.y. + 1 km.$ At the center of the shell is a transparent, regulation-size billiard ball with zero charge and zero angular momentum. The point of view of this description is an inertial frame in which the billiard ball and the coincident center of the shell are at rest. The billiard ball is naturally the repository and embodiment of all its own inherent information (e.g., the relative positions and states of its constituent atoms).  The mass of the billiard ball is considered to be too small to significantly affect the surrounding space-time curvature. This configuration is unchanged for thousands of years. During all that time the space between the billiard ball and the shell has a flat Minkowski metric. 

After those uneventful years the entire shell undergoes a sudden and uniform phase transition into dust that collapses radially toward the center. Applying Birkhoff’s theorem \cite{2} for the Oppenheimer-Snyder collapse \cite{3} of a shell of dust we have a static Schwarzschild metric outside the shell (though the shell is collapsing) and a static, flat Minkowski metric between the billiard ball and the shell. (In general relativity, Birkhoff's theorem states that any spherically symmetric solution of the vacuum field equations must be static and asymptotically flat.)

\begin{figure}
\centering
\includegraphics[width=0.8\textwidth]{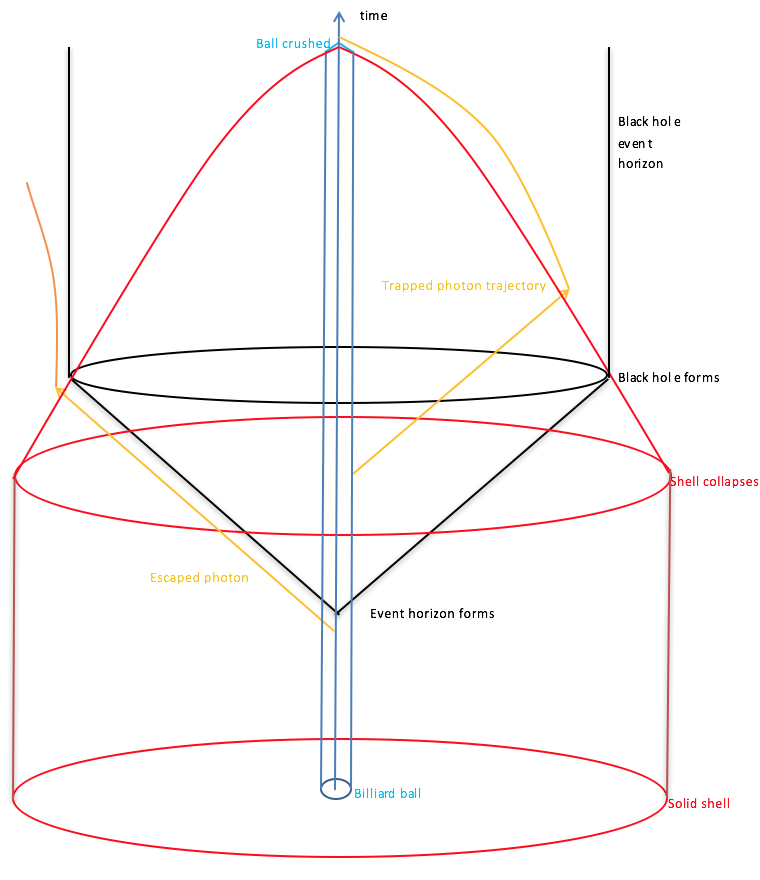}
\caption{\label{fig:solid-shell}solid shell collapsing on a billiard ball (one space dimension suppressed)}
\end{figure}

As soon as the shell of dust has travelled 1 km a BH forms. (See Figure: \ref{fig:solid-shell}) Henceforth, no information located inside the BH could ever get to the BH’s event horizon. Actually, the event horizon, from which no information can escape, starts forming 1 year before the BH forms. Indeed, photons emitted by the ball slightly less than 1 year before the BH forms will travel away from the ball for 1 year, they will then meet the freshly formed BH event horizon, and be trapped within the sphere it delimits. Photons emitted by the ball more than 1 year before the BH formed will be able to escape and could eventually be seen by a distant observer, though increasingly red shifted and dimmer.

All photons emitted by the ball less than 1 year before the BH formed will eventually rain down on to the billiard ball about 1 year after the BH formed and will then proceed to the BH singularity along with the crushed ball itself, as schematically illustrated in Figure 1. Notice that when the event horizon starts forming, during the year before the collapse of the large shell, there is no change or disturbance whatsoever (space time metric, forces) inside the shell and, in particular, not in the vicinity of the ball where the event horizon is forming.

Per standard General Relativity theory, the earliest possible time that the ball can have any inkling of the implosion of the shell, such as through any change in the local space time metric, is 1 year \emph{after} the formation of the BH. At that point no information from the ball could possibly move away from the ball, let alone reach the BH horizon 1 l.y. away.  Since none of the information embodied in the ball after the formation of the event horizon can ever reach the BH horizon it could never be encoded in any physical manner just outside the BH horizon. As the BH evaporates and disappears the information inside the ball vanishes forever, apparently violating unitarity and conservation of information.  In particular, a universe-filling space-like hypersurface (i.e., a Cauchy surface) including the billiard ball with its internal information at a time one thousand years prior to the formation of the BH will embody information about the state of the interior of the ball that is not available to a universe-filling space-like hypersurface a long time after the BH has evaporated.

BHC attempts to save this information from oblivion by postulating that the event horizon is enveloped by an extended horizon, about one Planck unit larger than the event horizon itself, and that all the information that would have been inside the event horizon is encoded in this extended horizon. Thus, this extended horizon must be initiated at the same time as the event horizon, which in the current case is one year before the formation of the BH that causes it to be brought into existence. It then expands at the speed of light along with the event horizon until the latter encounters the collapsing shell. All the information initially in the billiard ball, per BHC, somehow becomes encoded in this extended horizon as it expands through that billiard ball and therefore is available to the subsequent Hawking radiation. Thus, while BHC provides a path for the billiard ball information to become available to the Hawking radiation released in the subsequent evaporation of the BH, this is accomplished at the expense of causality since a physical extended horizon is being initiated one year \emph{before} the BH creation event that actually causes the creation of that extended horizon.    

The acausality of BHC can be further illustrated with a related, still essentially gravitationally spherically symmetric, thought experiment (suggested by Andre Mirabelli\cite{4}). Instead of a solid shell, we now have a dust shell with the same characteristics as before but freely collapsing from rest from a radius of three light years.  An experimenter places a thin, light board, opaque (to dust), suspended from afar on a stationary axis at one l.y. and 1 km from the center of the shell in a plane that passes through the center of the shell. The board will not affect the passage of the in-falling dust and hence the formation of the stretched horizon. However, just before the collapsing shell reaches the board the experimenter may decide to flip the board so it becomes perpendicular to the line connecting its center to the center of the shell. Such action would delay some of the dust from falling in, thereby delaying the formation of the extended horizon. Thus, the choice implemented by the experimenter has an immediate effect on an event 1 l.y. away, in violation of causality.     

The issue of teleology and causality is recognized by Susskind et. al. \cite{1} which addresses it in part with: “From a formal point of view, the cause of the horizon expansion is a gravitational dressing which is attached to the incoming energy flux.”  However, in our scenario the horizon expansion starts in a flat, static environment, which has been unchanged for thousands of years, about one year before the shell starts to collapse and hence no incoming energy flux exists to carry a “gravitational dressing” which can be “the cause of the horizon expansion.” 

Some of the issues raised herein are in line with similar concerns expressed by Roger Penrose\cite{6} regarding information conservation and teleological aspects of the event horizon. Notice, however, that Penrose (along with Susskind et. al.\cite{1} ) focus on matter falling into an existing BH horizon whereas here we consider matter engulfed by an expanding horizon. One might attempt to argue that there is no difference between matter falling into a horizon or being engulfed by an expanding horizon. There is one crucial difference tough: in the case of matter falling into a BH horizon the metric inside the horizon is highly curved, with the light cone tilted toward the singularity; in our case the metric inside the expanding horizon, as well as just outside the horizon is a uniformly flat Minkowski metric. So, the extreme curvature that is available in the case of matter falling into an existing BH, which can be the situation that can initiate the transfer of any information falling into a BH to a proposed membrane, is not available to initiate such a transfer in the billiard ball thought experiment described in this paper.

As noted by John H. Conway and Simon Kochen\cite{5}: "It is usual tacitly to assume the temporal causality principle that the future cannot alter the past. Its relativistic form is that an event cannot be influenced by what happens later in any given inertial frame.” Therefore, the premise that BHC requires that the instantiation of the physical extended horizon is triggered by the formation of the BH 1 year in the future, explicitly violates causality. 

The next section analyzes a slightly different scenario and shows that while BHC saves BH information from destruction, the acausality it engenders allows for the controllable acausal transmission of information from the future to the present.

\section{Massive solid sphere and thin shell}

Consider a solid, homogenous, stationary, electrically neutral, massive sphere of mass M such that $2GM/c2 = 1 l.y. - 1 km.$ The sphere is $100\%$ transparent and has a refractive index equal to one at all wave lengths. It has a radius $R=1 l.y. - 0.5 km.$ The sphere is at the center of a uniform, spherical shell, made of dust with the same optical properties as the sphere and has a mass m such that $2Gm/c2 = 1 km$. The shell has a radius much larger than 1 l.y. and is collapsing into the sphere. When the radius of the shell reaches 1 l.y. a BH forms. As the shell continues its collapse the sphere collapses along with the shell into a singularity.

The BH event horizon starts forming at the center of the sphere at time T (computed below) before the BH forms, and its radius expends at the speed of light (subject to space-time curvature, see below) from a Planck length to 1 l.y. (Schematically illustrated in Figure 2.) According to BHC the event horizon is surrounded by a stretched horizon, 1 Planck unit larger, that encodes all the information inside the horizon. As the horizon expands at the speed of light, so does the information-carrying stretched horizon. Therefore, the information can only be encoded by massless particles. For simplicity, and without loss of generality, we assume that the information is encoded entirely by photons; we thus ignore the possibility that the information may be encoded by gravitons, dark photons, or any other massless particles. As the horizon reaches its ultimate size (ignoring any subsequent Hawking radiation) of 1 l.y. the stretched horizon consists of photons about 1 Planck length outside the BH horizon and moving radially away from the BH. The gravitational pull outside a 1 l.y. BH is quite moderate, the geometry outside that BH is rather flat, and the photons continue their outward journey with only moderate delay and red shifting.  Bob, a stationary distant observer will therefore notice a single flash of light emitted when the BH is formed and encoding all the information initially in the sphere. Therefore, immediately after the formation of the BH the ensuing Hawking radiation would be almost content-free thermal radiation, except for information in the thin shell and stuff falling into BH after its formation, rather than engulfed by the expanding extended horizon.   
To compute the time T it takes a flash of light to travel from the center of the sphere to its boundary, or equivalently, indicating how long before the formation of the BH the BH event horizon started to form, we start with the Schwarzschild metric
\[
ds^{2}={\left(1-\frac{2MG}{r}\right)}{dt}^2-{\left(1-\frac{2MG}{r}\right)}^{-1}{dr}^2-r^2d{\theta{}}^2-r^2{sin}^2\theta{}d{\varphi{}}^2
\]
in units where the speed of light in vacuum $c=1$, $t$ is the Schwarzschild time of a standard clock at rest at spatial infinity, and the Schwarzschild radial coordinate $r$ is defined so that the area of the 2-sphere at $r$ is $4\pi r^2$.

For photons flowing radially from the origin we have:
\[
ds =\ d{\theta{}} =\ d{\varphi{}} =0
\]
Thus, the time $T$ to reach a distance R from the origin is given by:
\[
T=\int_0^Tdt=\int_0^R\frac{dr}{\left(1-\frac{2M(r)G}{r}\right)}
\]

Where M(r) is the mass inside a sphere of radius $r$ and is given by:
$$
M(r)=\frac{4}{3}\pi\rho r^{3}
$$
And where $\rho$ is the uniform density of the sphere. Performing the integral we get:
$$
T=\frac{1}{2p}\left[\log(1+pR) - \log(1-pR)\right]
$$
Where
$$
p=\sqrt{\frac{8G\pi\rho}{3}}
$$
Thus, unsurprisingly, $T\rightarrow\infty$ as $R\rightarrow 1/p$ or as the density of the sphere increases to the point where the sphere itself, without the addition of the dust shell becomes a BH. Indeed, when M is such that $R=2GM$ we have
$$
R=2G\frac{4}{3}\pi\rho R^3
$$
Thus, giving $R=1/p$ and leading to $T=\infty$. 

By adjusting the radius and density of the sphere we can control $T$ to be any value larger than $R$.   With this flexibility it is possible (in the realm of thought experiments) to design a whimsical contraption that allows asking a question about the outcome of an event at some pre-determined time in the future and receiving the answer in the present. The contraption may have a radius of, e.g., about 1 km and weigh about $1/3$ solar mass. While it turns out (as shown below) that the device is useless for profitable prognostications, it does prove handy for some theoretical analyses. 

The device consists of a solid sphere at the center of a solid spherical shell, similar to the setup above. The radii and masses of the sphere and the shell are fine-tuned so that when the shell collapses and forms a BH with the sphere the corresponding event horizon and associated 
stretched horizon will start forming at time $T$ before the BH forms.  Inside the sphere, at say about 1 mm from the center of the sphere there is an observer, Alice, equipped with a photodetector focused on the center, and committed to broadcast a signal to the outside if at some specified time (the present, in our scenario) she has not detected any light from the center. The shell is under the control of Bob, committed to ascertain if some specific event (e.g., the temperature in Central Park is at least 30 degrees) occurs at time T in the future. If the event does occur at time T Bob triggers an immediate uniform phase transition of the shell from solid to dust so it collapses and forms a BH along with the sphere.  (The uniformity of the phase transition can be achieved with fine-tuned wiring, with delay elements as needed, to phase transition triggers distributed uniformly over the shell.)

\begin{figure}
\centering
\includegraphics[width=0.8\textwidth]{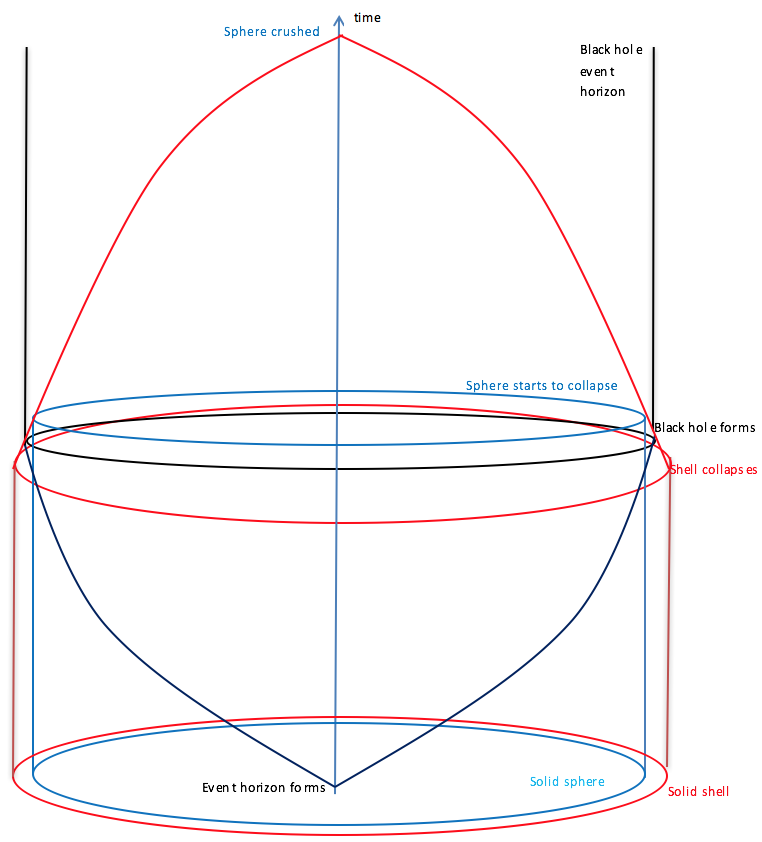}
\caption{\label{fig:solid-sphere}solid shell collapsing on a solid sphere (one space dimension suppressed)}
\end{figure}

Thus, if the watched for event does happen at time T in the future Alice’s photodetector will detect the light from the expanding stretched horizon at the present, and Alice will thereby be engulfed and doomed by the event horizon.  Alice will “know” that the watched for event happens at time T in the future, she will have a short time to savor this knowledge, but she will have no way of sharing this knowledge with anyone outside the event horizon.  If the watched for event at time T does not happen Bob does nothing, the shell remains solid, no BH is formed, no expanding event horizon and hence no stretched horizon materialize and Alice’s photodetector does not detect any light. Therefore, Alice knows now that the watched for event will not happen at time T in the future and broadcasts a signal announcing that at the present time she has learned that at future time T the watched for event does not happen. The travel time of the message through the sphere will be T and the anxious recipients of the message will therefore not get it until time T in the future, at which point they already know the outcome of the event. 

Alice receiving information by not detecting any flash of light is analogous to a null measurement such as the negative-result experiment described by Renninger \cite{7}. 
While the apparatus described above, even if it could be built, is useless for receiving information from the future in the present outside that apparatus, it is nevertheless clear that Alice does learn now about the outcome of a future event, takes specific action based on this knowledge, and Bob can ascertain that indeed she actually did act upon the advanced knowledge, in violation of causality.  The Appendix explores the ramifications of this scenario in more detail.  

\section*{Discussion}

This paper argues that the initial formulation of BHC violates causality, in particular it allows for the transfer of information from the future to the present and, as shown in the Appendix, can lead to internal inconsistencies. Those arguments, however, may not applicable to more recent developments such as fuzzballs\cite{8}.  Nevertheless, the scenarios presented herein might be customized to analyze other proposed solutions to the information paradox\cite{9} to see which ones preserve causality.

\section*{Acknowledgements}
I have benefited from valuable discussions with members of the NY Quantum Theory Group, in particular, many comments and enlightening conversations with Dr. Andre Mirabelli.

\section*{Appendix} 

This Appendix exploits the contraption described at the end of Section 3 to present a more picturesque yet more compelling scenario for causality violation; it then generalizes the procedure to show that BHC allows the transmission of unlimited amount of information from the future to one recipient in the present.

Alice and Charlie are both terminally ill and have only 3 months to live. While incurable they both retain all their faculties without experiencing any discomfort. They each have an earth shaking secret. The secrets are unrelated and no one else knows that those secrets exist, let alone that Alice and Charlie are in possession of those secrets. Alice has decided not to expose her secret in any way unless she is sure that some event E occurs 20 years in the future; e.g., the temperature in Central Park at noon 2037-1-1 is at least 30 degrees. Charlie is equally determined to keep his secret unless event E does not occur 20 years in the future; e.g., the temperature in Central Park at noon on 2037-1-1 is below 30 degrees. In order to accommodate both secret holders each is provided with a contraption as described in Section 3. Bob is committed to trigger the collapse of Alice’s (Charlie’s) shell if the watched for event E at time $T$ (noon on 2037-1-1) does not (does) happen. 

The following paragraph provides a textual narrative of the subsequently unfolding events. Table \ref{tab:causal}, however, may be a more intuitive presentation of those events; it provides a causal flow (the events in each row can only affect the rows below) which is strikingly different from the temporal chronology denoted in the left column of the table.  The time ticks that everyone uses are provided by a standard clock at rest at spatial infinity (i.e., the Schwarzschild time).

Alice (Charlie) is determined to broadcast her (his) secret as well as announcing that at the present time she (he) has learned that at future time T the watched for event E does (does not) happen, if at the present she (he) does not detect any light from the center of her (his) sphere. Thus, if at time T in the future the watched for event E does (does not) happen Bob will trigger the collapse of Charlie’s (Alice’s) shell; he will immediately receive a message from Alice (Charlie) declaring that at the present she (he) knew that the event at time T in the future does (does not) happen along with Alice’s (Charlie’s) totally unexpected and utterly surprising secret. 

As soon as Bob triggers the collapse of Charlie’s (Alice’s) shell Bob actually sees Alice (Charlie) confidently broadcasting her (his) secret message.

\begin{table}
\centering 
\begin{tabular}{|p{49pt}|p{193pt}|p{197pt}|}
\hline
\parbox{49pt}{\centering 
{\small Time}
} & \multicolumn{2}{|c|}{\parbox{390pt}{\centering 
{\small Events}
}} \\
\hline
\parbox{49pt}{\raggedright 
{\small 2016-12-31 12:00:00''}
} & \multicolumn{2}{|l|}{\parbox{390pt}{\raggedright 
{\small Alice and Charlie enter their respective contraptions.}
}} \\
\hline
\parbox{49pt}{\raggedright 
{\small 2016-12-31 12:00:00''}
} & \multicolumn{2}{|l|}{\parbox{390pt}{\raggedright 
{\small Bob commits to trigger the collapse of Charlie's shell if event E occurs
on 2037-1-1 at noon and to trigger the collapse of Alice's shell if event E does
not occurs on 2037-1-1 at noon. }
}} \\
\hline
\parbox{49pt}{\raggedright 
{\small 2037-1-1 12:00:00''}
} & \parbox{193pt}{\raggedright 
{\small Event E occurs on 2037-1-1 at noon.}
} & \parbox{197pt}{\raggedright 
{\small Event E does not occurs on 2037-1-1 at noon. }
} \\
\hline
\parbox{49pt}{\raggedright 
{\small 2037-1-1 12:00:01''}
} & \parbox{193pt}{\raggedright 
{\small Bob triggers the collapse of Charlie's shell. }
} & \parbox{197pt}{\raggedright 
{\small Bob triggers the collapse of Alice's shell.}
} \\
\hline
\parbox{49pt}{\raggedright 
{\small 2037-1-1 12:00:02''}
} & \parbox{193pt}{\raggedright 
{\small Charlie's shell starts to collapse. }
} & \parbox{197pt}{\raggedright 
{\small Alice's shell starts to collapse.}
} \\
\hline
\parbox{49pt}{\raggedright 
{\small 2037-1-1 12:00:03''}
} & \parbox{193pt}{\raggedright 
{\small Charlie's shell and sphere are now a BH.}
} & \parbox{197pt}{\raggedright 
{\small Alice's shell and sphere are now a BH.}
} \\
\hline
\parbox{49pt}{\raggedright 
{\small 2017-1-1 12:00:03''}
} & \parbox{193pt}{\raggedright 
{\small An extended horizon starts forming at the center of Charlie's sphere and
expands at the speed of light.}
} & \parbox{197pt}{\raggedright 
{\small An extended horizon starts forming at the center of Alice's sphere and
expands at the speed of light.}
} \\
\hline
\parbox{49pt}{\raggedright 
{\small 2017-1-1 12:00:03'' + 1 nanosecond}
} & \parbox{193pt}{\raggedright 
{\small Charlie sees the expanding extended horizon as a flash of light,
realizes that his condition for exposing his secret is not fulfilled and lives
quietly his remaining days inside the newly formed event horizon.  Alice does not
see any flash of light from the center of her sphere, she now knows that event E
does happen in 2037-1-1 12:00:00'' and starts broadcasting her secret.}
} & \parbox{197pt}{\raggedright 
{\small Alice sees the expanding extended horizon as a flash of light, realizes
that her condition for exposing her secret is not fulfilled and lives quietly her
remaining days inside the newly formed event horizon. Charlie does not see any
flash of light from the center of his sphere, he now knows that event E does
happen in 2037-1-1 12:00:00'' and starts broadcasting his secret.}
} \\
\hline
\parbox{49pt}{\raggedright 
{\small 2017-1-1 12:15:00''}
} & \parbox{193pt}{\raggedright 
{\small Alice finishes broadcasting her secret.}
} & \parbox{197pt}{\raggedright 
{\small  Charlie finishes broadcasting his secret.}
} \\
\hline
\parbox{49pt}{\raggedright 
{\small 2017-3-31 12:00:00''}
} & \multicolumn{2}{|l|}{\parbox{390pt}{\raggedright 
{\small Alice and Charlie die peacefully in their sleep.}
}} \\
\hline
\parbox{49pt}{\raggedright 
{\small 2037-1-1 12:00:04''}
} & \parbox{193pt}{\raggedright 
{\small Bob, and the rest of the world, start receiving Alice's unexpected and
astonishing secret as it emerges from Alice's contraption. They also see a flash
of light from Charlie's contraption as it turns into a BH and its extended
horizon escapes the shell.}
} & \parbox{197pt}{\raggedright 
{\small Bob, and the rest of the world, start receiving Charlie's unexpected and
astonishing secret as it emerges from Charlie's contraption. They also see a
flash of right from Alice's contraption as it turns into a BH and its extended
horizon escapes the shell.}
} \\
\hline
\parbox{49pt}{\raggedright 
{\small 2037-1-1 12:15:04''}
} & \parbox{193pt}{\raggedright 
{\small Bob, and the rest of the wold, finish receiving Alice's secret as it
emerges from Alice's contraption. The new-found knowledge changes the course of
history.}
} & \parbox{197pt}{\raggedright 
{\small Bob, and the rest of the world, finish receiving Charlie's secret as it
emerges from Charlie's contraption. The new-found knowledge changes the course hf
history.}
} \\
\hline
\parbox{49pt}{\raggedright 
{\small 2037-1-1 \textasciitilde{}12:20:00''}
} & \parbox{193pt}{\raggedright 
{\small Charlie's contraption, along with Charlie's remains collapse into a
singularity.  }
} & \parbox{197pt}{\raggedright 
{\small Alice's contraption, along with Alice's remains collapse into a
singularity. }
} \\
\hline
\end{tabular}
 
\caption{\label{tab:causal}Causal flow (top to bottom) of backwards information transmission example.}
\end{table}

Thus, Alice and Charlie are acting at present based on a future event which takes place long after they are both dead; furthermore, their respective actions, i.e., which secret is revealed will affect differently subsequent events. 

As can be seen from the Table Bob does not see any violation of causality. The significance of this observation is made explicit by allowing Alice and Charlie to enter their respective contraptions in company of their respective inseparable friends Debbie and Ed. Debbie and Ed take detailed notes of everything they observe. As soon as Alice and Charlie die Debbie and Ed begin the long journey out of the contraption. Of course, only one of them, say Debbie, will succeed since the other’s contraption collapses into a singularity. Debbie can then compare notes with Bob, they quickly realize that their respective observations – Debbie witnessed violation of causality while Bob saw no such thing – are mutually exclusive; arguably even more severe outcome than violation of causality.

\begin{figure}
\centering
\includegraphics[width=0.8\textwidth]{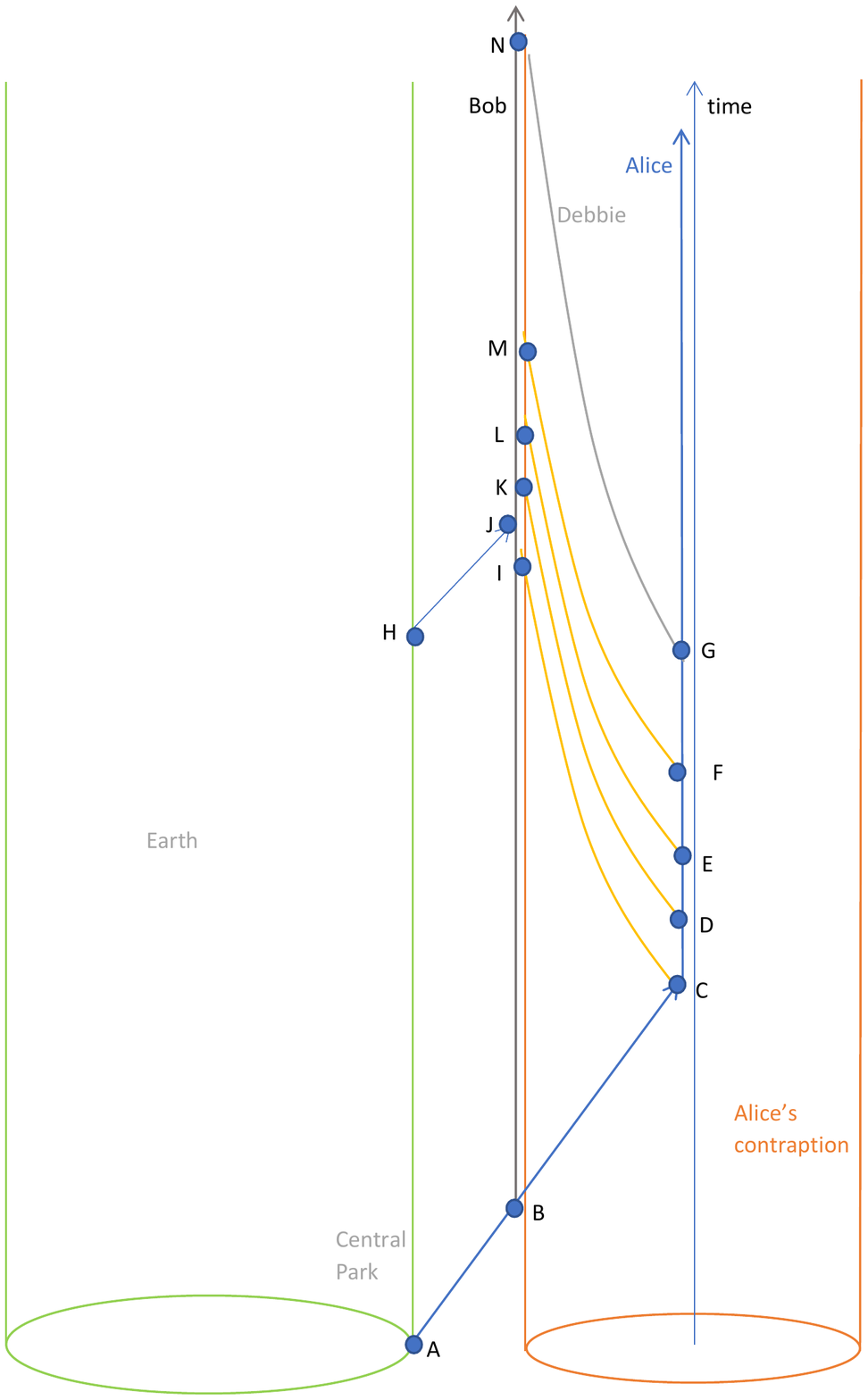}
\caption{\label{fig:a_1}Alice’s story (one space dimension suppressed))}
\end{figure}

Figure \ref{fig:a_1} provides a schematic (i.e., very much not to scale) depiction of a simplified version of the scenario discussed above. Charlie is completely omitted from the picture and Bob stays right next to Alice’s contraption. In this picture Earth may be in circular orbit of perhaps 100,000 km around the much heavier contraption. The events in the figure are labeled alphabetically (A, B, … N) in chronological order (2016-12-31 10:00:00” to 2057-3-31 12:10:00”) . The key to those events is provided in table \ref{tab:a_2}. Just as table \ref{tab:causal} the events in table \ref{tab:a_2} are listed in causal, rather than chronological order.

The preceding scenario can be generalized to transmit N bits from the future to the present. This is accomplished with S secrets holders where
$$
S=2^{N}
$$
Each secrets holder enters a contraption as described in Section 3 and has an ordered set of N secrets. The $S\times N$ secrets are unrelated. Each secrets holder further has a unique array of N bits corresponding to the occurrence or non-occurrence of N events at specified times in the future.
Bob has a copy of each of the N arrays. If the kth event does (does not) happen at the kth specified time then Bob does (does not) trigger the collapse of the shells of the secrets holders with 0 (1) in the kth position in their respective arrays. Each event has an a {\em priori} probability of $50\%$ of occurring at its specified time.
Each of the secrets holders broadcasts their kth secret if each of the first k events did (did not) happen if the array of the secrets holder has 1 (0) in the corresponding position.

The number of secrets holders increases exponentially with the number of bits to be transmitted from the future to the present. Each secrets holder has one of the S possible arrays of N bits and one of them is notified at present as the winner of the future lottery.  Half of the secrets holders never broadcast a single secret. Only one broadcasts a full set of N secrets. While this procedure is highly inefficient by modern communications standards, it demonstrates a clear violation of causality nonetheless.

\begin{table}
\centering 
\begin{tabular}{|p{37pt}|p{51pt}|p{160pt}|p{161pt}|}
\hline
\parbox{37pt}{\raggedright 
{ Event identifier}
} & \parbox{51pt}{\raggedright 
{ Time}
} & \parbox{160pt}{\raggedright 
{ Alice, Debbie}
} & \parbox{161pt}{\raggedright 
{ Bob}
} \\
\hline
\parbox{37pt}{\raggedright 
{ A}
} & \parbox{51pt}{\raggedright 
{ 2016-12-31 10:00:00''}
} & \multicolumn{2}{|l|}{\parbox{322pt}{\raggedright 
{ Alice, Bob and Debbie leave Earth for Alice's contraption. Bob commits
to trigger The collapse of Alice's shell if agreed event does not occur on
2037-1-1 at noon.}
}} \\
\hline
\parbox{37pt}{\raggedright 
{\small B}
} & \parbox{51pt}{\raggedright 
{\small 2016-12-31 12:00:00''}
} & \parbox{160pt}{\raggedright 
{\small Alice and Debbie enter Alice's contraption.}
} & \parbox{161pt}{\raggedright 
{\small Bob remains just outside Alice's contraption.}
} \\
\hline
\parbox{37pt}{\raggedright 
{\small C}
} & \parbox{51pt}{\raggedright 
{ 2016-12-31 12:01:00''}
} & \parbox{160pt}{\raggedright 
{ Alice and Debbie arrive to their post at Alice's contraption.}
} & \parbox{161pt}{\raggedright } \\
\hline
\parbox{37pt}{\raggedright 
{ I}
} & \parbox{51pt}{\raggedright 
{ 2036-12-31 12:01:00''}
} & \parbox{160pt}{\raggedright } & \parbox{161pt}{\raggedright 
{ Bob sees Alice and Debbie arrive to their post at Alice's concentration.}
} \\
\hline
\parbox{37pt}{\raggedright 
{ H}
} & \parbox{51pt}{\raggedright 
{ 2037-1-1 12:00:00''}
} & \parbox{160pt}{\raggedright } & \parbox{161pt}{\raggedright 
{ The Weather Channel broadcasts that the temperature in Central Park at
noon 2037-1-1 is 40 degrees.}
} \\
\hline
\parbox{37pt}{\raggedright 
{ J}
} & \parbox{51pt}{\raggedright 
{ 2037-1-1 12:03:00''}
} & \parbox{160pt}{\raggedright } & \parbox{161pt}{\raggedright 
{ Bob receives the Weather Channel broadcast that the temperature en
Central Park at noon 2037-1-1 is 40 degrees and decides not to trigger the
collapse of Alice's shell.}
} \\
\hline
\parbox{37pt}{\raggedright 
{ D}
} & \parbox{51pt}{\raggedright 
{ 2017-1-1 12:03:01''}
} & \parbox{160pt}{\raggedright 
{ Alice does not see any flash of light from the center or her sphere; she
now knows that the temperature in Central rank at noon 2037-1-1 is above 30
degrees and starts broadcasting the secret.}
} & \parbox{161pt}{\raggedright } \\
\hline
\parbox{37pt}{\raggedright 
{ K}
} & \parbox{51pt}{\raggedright 
{ 2037-1-1 12:03:01''}
} & \parbox{160pt}{\raggedright } & \parbox{161pt}{\raggedright 
{ Bob starts receiving Alice's message.}
} \\
\hline
\parbox{37pt}{\raggedright 
{ E}
} & \parbox{51pt}{\raggedright 
{ 2017-1-1 12:18:01''}
} & \parbox{160pt}{\raggedright 
{ Alice finishes transmitting her message}
} & \parbox{161pt}{\raggedright } \\
\hline
\parbox{37pt}{\raggedright 
{ L}
} & \parbox{51pt}{\raggedright 
{ 2037-1-1 12:18:01''}
} & \parbox{160pt}{\raggedright } & \parbox{161pt}{\raggedright 
{ Bob vanishes receiving Alice's message.}
} \\
\hline
\parbox{37pt}{\raggedright 
{ F}
} & \parbox{51pt}{\raggedright 
{ 2017-3-31 12:00:00''}
} & \parbox{160pt}{\raggedright 
{ Alice dies peacefully in her sleep.}
} & \parbox{161pt}{\raggedright } \\
\hline
\parbox{37pt}{\raggedright 
{ M}
} & \parbox{51pt}{\raggedright 
{ 2037-3-31 12:00:00''}
} & \parbox{160pt}{\raggedright } & \parbox{161pt}{\raggedright 
{ Bob sees Alice dying peacefully in her sleep.}
} \\
\hline
\parbox{37pt}{\raggedright 
{ G}
} & \parbox{51pt}{\raggedright 
{ 2017-3-31 12:10:00''}
} & \parbox{160pt}{\raggedright 
{ Debbie starts the journey outside the sphere; with a hugely powerful
vehicle the journey lasts 40 years.}
} & \parbox{161pt}{\raggedright } \\
\hline
\parbox{37pt}{\raggedright 
{ N}
} & \parbox{51pt}{\raggedright 
{ 2057-3-31 12:10:00''}
} & \multicolumn{2}{|l|}{\parbox{322pt}{\raggedright 
{ Debbie emerges from Alice's contraption and compares notes with Bob
about their respective observation regarding Alice. Debbie's notes show a clear
violation of causality while Bob's notes record perfect agreement with
causality.}
}} \\
\hline
\end{tabular} 
\caption{\label{tab:a_2}Events in Alice’s story}
\end{table} 

\end{document}